# Self-tearing and self-peeling of folded graphene nanoribbons

*Alexandre F. Fonseca\* and Douglas S. Galvão\**

Applied Physics Department, State University of Campinas, Campinas, SP, 13083-970, Brazil.

**\* Corresponding Authors**

Tel: +55 19 35215364. E-mail: afonseca@ifi.unicamp.br (Alexandre F. Fonseca);

Tel: +55 19 35215373. E-mail: galvao@ifi.unicamp.br (Douglas S. Galvão).

**ABSTRACT:** A recent experimental study showed that an induced folded flap of graphene can spontaneously drive itself its tearing and peeling off a substrate, thus producing long, micrometer sized, regular trapezoidal-shaped folded graphene nanoribbons. As long as the size of the graphene flaps is above a threshold value, the "tug of war" between the forces of adhesion of graphene-graphene and graphene-substrate, flexural strain of folded region and carbon-carbon (C-C) covalent bonds favor the self-tearing and self-peeling off process. As the detailed information regarding the atomic scale mechanism involved in the process remains not fully understood, we carried out atomistic reactive molecular dynamics simulations to address some features of the process. We show that large thermal fluctuations can prevent the process by increasing the probability of chemical reactions between carbon dangling bonds of adjacent graphene layers. The effects of the strength of attraction between graphene and the substrate on the ribbon growth velocities at the early stages of the phenomenon were also investigated. Structures with initial armchair crack-edges were observed to form more uniform cuts than those having initial zigzag ones. Our results are of importance to help set up new experiments on this phenomenon, especially with samples with nanoscale sized cuts.

# 1. Introduction

The advent of graphene created a new revolution in materials science. Graphene presents unique electronic [1], thermal [2] and mechanical [3] properties. Recent advances on large-scale syntheses [4,5], create new perspectives for applications in several areas [6,7].

Recently, an experiment with graphene membranes revealed a new, simple but surprising, phenomenon: the spontaneous self-tearing and self-peeling off towards the formation of a folded micrometer-size graphene nanoribbon [8]. This experiment consists of after making a hole in graphene, it induces the formation of folded ribbons on the sides of the hole. The folded nanoribbon flaps spontaneously pull, tear and peel off the behind graphene structure so driving their own propagation. If the initial cuts produce a hole with a polygonal shape of *n* sides, the final structure would look like a flower of *n* petals[1], each petal being one folded graphene nanoribbon generated by itself. Experimental results were analyzed in terms of strain energy of folded region, energy necessary to break carbon-carbon (C-C) bonds and interplay of adhesion energies between graphene-substrate and graphene-graphene layers. The most important characteristics of this process are: i) micrometer size of the folded ribbon and minimum ribbon width size to maintain the self-propagating process; ii) increasing temperature enhancement of the process; iii) period of time of the process of the order of few days or about nanometer per second of magnitude of nanoribbon propagation velocity. As from the experiment we cannot determine the initial steps of the growth/propagation, as well as, the atomistic details of the tearing and peeling off processes, several questions remain unclear and atomistic simulations might be helpful in order to provide insights on the mechanisms of the ribbon growth/propagation.

---

[1] See comments in: https://www.newscientist.com/article/2097359-graphene-sheets-open-like-a-flowers-petals-when-poked/ Accessed in 05-18-2018.

Recently, we have presented some results of molecular dynamics (MD) simulations of self-driven graphene tearing and peeling off the substrate [9], where the dynamics of folded graphene nanoribbons of different widths and two types of chiral cut edges were observed. We have obtained an estimate of the minimum width, $w$, of the folded graphene nanoribbon to initiate the nanoribbon growth as $w \sim 80$ nm, and we have estimated the values of the initial velocity of the ribbon front, at least at the onset of growth, from 1 to 5 m/s. These results are consistent with the experimental ones regarding the effects of decreasing width on crack propagation features. We have also observed not only positive but also negative effects of thermal fluctuations on the process: they, at the same time, can help promote the crack propagation by triggering the C-C bond breaking, as well as to prevent it by promoting some C-C re-bonding between the edges of the graphene layers. In this work we present an expanded and detailed investigation of the process. We investigated the local structure of the crack-edges of the systems, some properties of the nanoribbon growth process at different adhesion strengths between substrate and graphene, and the behavior of systems with width above the threshold size of 80 nm: $w \cong 160$ nm. The results obtained here can be summarized as: first, the minimum strength from which the initial velocity of the crack propagation starts being dependent on substrate adhesion strength is 0.03 J/m$^2$. Second, the shape of the as grown crack-edge from the structure with an initial armchair crack-edge is quite regular and presents zigzag pattern, while for the initial zigzag crack-edge it is not uniform. Third, the formation of C-C bonds between carbon atoms of different graphene layers are confirmed to occur for large temperatures and show to drastically reduce the nanoribbon growth/propagation velocity. Moiré-like patterns were observed during the growth/propagation process. Although the nanoribbon front edge becomes curved during the growth/propagation process, when the maximum growth occurs, the front edge becomes straight, independently of the initial shape of the crack-edge and at the cost of additional C-C bond breaks. Finally, we discuss the main features that can determine the ability to sustain the self-tearing of graphene and the corresponding nanoribbon growth/propagation.

There are many studies in the literature about the different conditions and external features concerning the peeling off process [10]. However, few studies analyzed a common feature observed when trying to peel off rectangular flaps: that they came out in small size non-rectangular shapes. Hamm *et al*. [11] showed that interplay between bending, fracture and surface energies explains the triangular narrowing and final collapse of peeled off thin films. Ibarra *et al*. [12] analyzed the crack propagation path in a teared/fractured sheet having anisotropies. Sen *et al*. [13] performed an atomistic study of the tearing and peeling off graphene sheets from adhesive substrates. They showed that, differently from the macroscopic mechanical analysis, the final geometry of the peeled off graphene depends on graphene-substrate strength of adhesion and number of layers. Huang *et al*. [14] also applied computational atomistic methods based on reactive MD force fields to investigate how to control tearing paths in graphene through chemical functionalization. Also using MD simulations, Wang and Liu [15] investigated the fracture toughness of graphene with grain boundaries, showing that these defects can block crack propagation. Recently, He *et al*. [16] studied the mechanics of graphene tearing and folding under the action of external forces or thermal fluctuations. Using MD simulations, they have found out scaling laws between the taper angle of the cut graphene nanoribbon and adhesive energy. They also showed that the low value of the bending stiffness allows for the stretching energy to play an important role to the maintenance of the graphene nanoribbon growth/propagation process. In fact, the low bending stiffness allows for the folded part of graphene to be easily formed at low radius of curvature [17]. There are many studies about rupture and crack propagation in graphene [18-26], but to our knowledge, there are no studies other than those of References [9, 11-16] on the crack propagation in 2D materials specifically through peeling off-like processes.

This work is organized as follows. In Section **2**, we describe the system models and computational methods employed in this work. In Section **3**, the results are presented and discussed. In Section **4**, we summarize the results and present the conclusions.

## 2. System model and computational methods

*2.1 System model*

The experiments on the self-driven growth of folded graphene nanoribbons were performed on micrometer sized samples [8]. Even using classical molecular dynamics (MD) we cannot simulate such an order of magnitude in size. In order to simulate a system as close as possible to the experimental setup, we devised a system at nanoscale with just one petal or folded graphene nanoribbon and fixed extremities in order to mimic the inertial effect of large systems. We considered only one graphene layer on top of a substrate. In Figure **1**, we present the details of the overall shape of the structures and two chiral models of an initial folded graphene nanoribbon of initial width *w*.

The values of initial width, *w*, considered in this study are ~ 80 and 160 nm. Detailed structural information are presented in Table **I**. One of the zigzag crack-edge systems of nanoribbon width *w* = 80 nm, possesses the largest length along *x* direction (see Figure **2**). The system was chosen in order to determine the dependence of the local atomistic aspects of the crack propagation process with the system size during the growth/propagation. We have reported before [9] some initial results for the system having the same nanoribbon width, *w* = 80 nm, and zigzag crack-edge along the *x* direction, but of an overall size of 20 nm along *x* direction. The results obtained for the structure and dynamics of crack propagation of the structure in Ref. [9] are quantitatively and qualitatively very close to those reported here for the larger size system. In Table **I**, the *x* direction is defined as that of nanoribbon growth.

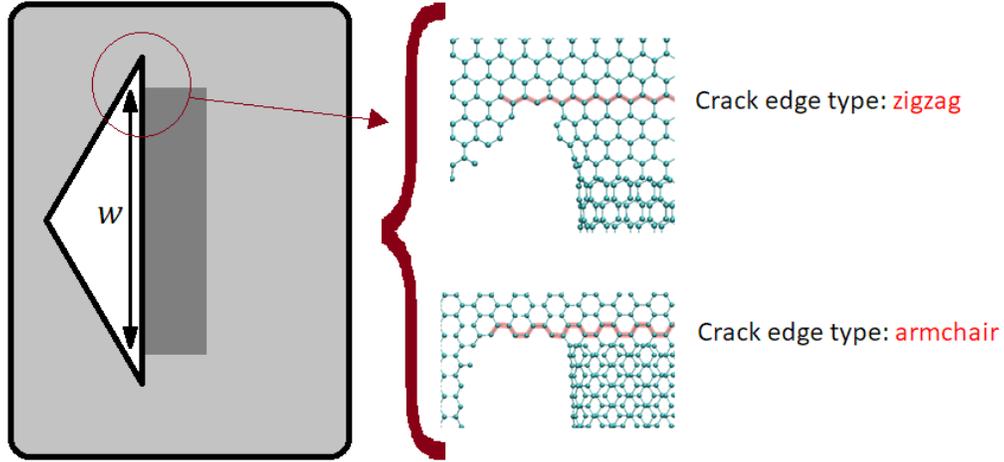

**Figure 1**: Scheme of the system model of a large piece of graphene sheet (light gray) with a triangular cut and one graphene nanoribbon already folded over the graphene sheet (dark gray). In the inset, the two types of modeled cut edges: zigzag and armchair. $w$ is the initial ribbon width. Pink lines are drawn to emphasize the edge chirality.

**TABLE I**. Structural information about the systems simulated here. $w$ is the nanoribbon initial width. $L_x$ and $L_y$ are the approximated values of the simulation box sizes along planar directions $x$ and $y$, respectively. For the crack-edge types, see Figure **1**.

| $w$ [nm] | ~ 80 | | ~ 160 | |
|---|---|---|---|---|
| **Crack-edge type** | zigzag | armchair | zigzag | Armchair |
| $L_x \times L_y$ [nm x nm] | 80 × 100 / 20 × 100 | 20 × 100 | 20 × 177 | 20 × 177 |
| # of carbon atoms | 307408 / 75184 | 82422 | 141090 | 145446 |

The substrate over which the graphene structure is deposited on is modeled by a 12/6 Lennard Jones potential, as given by:

$$E = 4\varepsilon \left[ (\sigma/r)^{12} - (\sigma/r)^{6} \right] \quad r < r_c, \qquad (1)$$

where $\varepsilon$, $\sigma$ and $r_c$ are the strength (in eV), the size of the particles as they interact with the wall, and the cutoff distance, respectively. $\sigma$ and $r_c$ are fixed in 3.5 Å and 12 Å, respectively, while $\varepsilon$ will be varied from 0.0005 to 0.01 eV, which correspond to adhesion strengths from 0.003 to 0.058 J/m$^2$.

*2.2 The force field and MD simulation protocols*

The force field used here is the AIREBO potential [27,28], that is available in LAMMPS computational package [29][2]. AIREBO is a well-known reactive force field used to study the structure and physical properties of many carbon nanostructures [30-36], including defective [37-39] and fractured graphene [16,19,22,26,40]. AIREBO abilities to describe quite well, both the structure and mechanical properties of graphene, as well as C-C bond breaking and bond formation, makes it a good choice for the study of self-tearing and self-folding of graphene nanoribbons. We used the correction of the initial cutoff distance of the potential to 2 Å in order to avoid the original overestimation of the breaking force of covalent C-C bonds, as usually made in other several studies of rupture of carbon bonds using REBO or AIREBO [26,31,40].

The protocols of the simulations try to mimic the experiment itself. The starting point of simulations is a cut graphene structure possessing an initial part of graphene structure folded over itself, as shown in Figure **1**. This structure is initially geometry optimized by conjugate gradient methods provided in LAMMPS package, with force tolerance of 10$^{-8}$ eV/Å. This initial step is made in order to obtain the structures in an energy minimum configuration before simulating the effects of thermal fluctuations. All the quantities that we are going to analyze as, for example, the front velocity, will be analyzed from this initial energy minimum configuration. A series of MD simulations at 300 K and 600 K using the Langevin thermostat are, then, performed for several nanoseconds, until the nanoribbon stops growing/propagation. MD time step and thermostat damping factors were set to 0.5 fs and 1 ps, respectively. In all cases, no periodic boundary conditions were used along the direction perpendicular to the graphene plane as allowed by the LAMMPS package. As mentioned before, for the dynamics simulations, the extremities of the system were kept fixed in order to mimic the inertial effects of a large system, as it is the case of the experiments [8]. Each series of MD simulations consists of running each structure, starting from the initial optimized configuration, by 2 ns at 300 K, then running it again by additional 5 ns at 600 K.

---

[2] LAMMPS Molecular Dynamics Simulator: http://lammps.sandia.gov/ Accessed on 05-18-2018.

By inspecting the structure as a function of time, it is possible to obtain an estimative for the average front velocity, local structure of the teared graphene nanoribbon, number of broken C-C bonds along the crack-edge, as well as the number of new C-C bonds formed among the graphene layers. The average front velocity will be estimated by calculating the change of the position of the middle part of the graphene nanoribbon front edge with respect to the position of the vertices of the triangular hole initially formed during the last 1 ns of simulation.

## 3. Results and Discussion

In Figure **2**, we present some representative MD snapshots of the test-structure investigated here with $w = 78.3$ nm and a zigzag crack-edge, simulated on a substrate with an adhesion strength $\varepsilon = 0.0005$ eV or 0.003 J/m$^2$. The Figure shows the evolution of the structure after the geometry optimization (top panel on the left). The second top panel from left to right of the Figure **2** shows the structure after 1 ns of simulation at 300 K. The front velocity during this first nanosecond is very high and several C-C bonds along the line of crack propagation broke as shown in the insets (carbon atoms from broken bonds are shown in pink color). Not shown in Figure **2** is the information that no additional break of C-C bonds along the cracking edge occurs during the second period of 1 nanosecond of simulation at 300 K, and that the front velocity decreased to 0.4 m/s. After that, we increased the temperature and after 1 ns at 600 K, new 11 C-C bonds at the crack edge broke. However, the front velocity remained at ~ 0.4 m/s. Additional 4 ns of simulation at 600 K were run and it was observed the breaking of only 5 more C-C bonds along the crack edge. Also, between the fourth and fifth nanosecond of simulation the formation of C-C bonds between the graphene layers (blue atoms in the inset of the bottom right panel of Figure **2**) was observed. The front velocity, then, decreased one order of magnitude to 0.02 m/s and we attribute this to the formation of the bonds between graphene layers, which prevents the growth/propagation process from continuing.

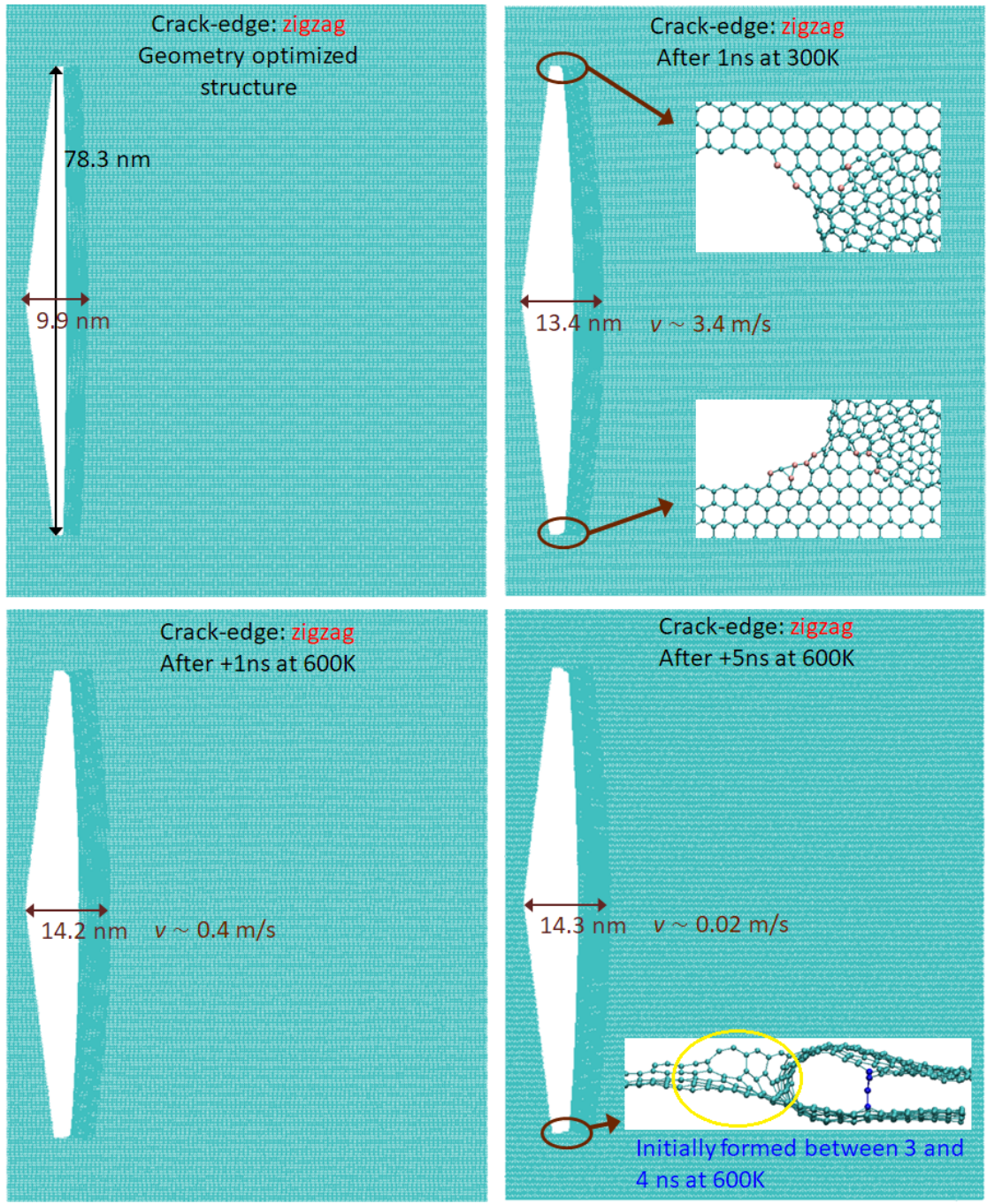

**Figure 2**: MD snapshots of the structure with nanoribbon width $w = 78.3$ nm and zigzag crack-edge simulated with a substrate with $\varepsilon = 0.0005$ eV or $0.003$ J/m². From top left to bottom right panels: geometry optimized, after 1 ns at 300 K, after 1ns at 600 K and after 5 ns at 600K, respectively. The horizontal arrow represents the distance from the vertices of the triangular cut to the nanoribbon front. The estimation of the front velocity is shown along the panels. The insets in the top right panel show the local structure of the crack-propagation where the atoms from the broken C-C bonds are presented in pink color. The inset in the bottom right panel highlights the formation of two C-C bonds (blue) between two adjacent graphene layers. In the last panel, the yellow circle highlights the out-of-plane segment of the graphene sheet after local crack-edge formation.

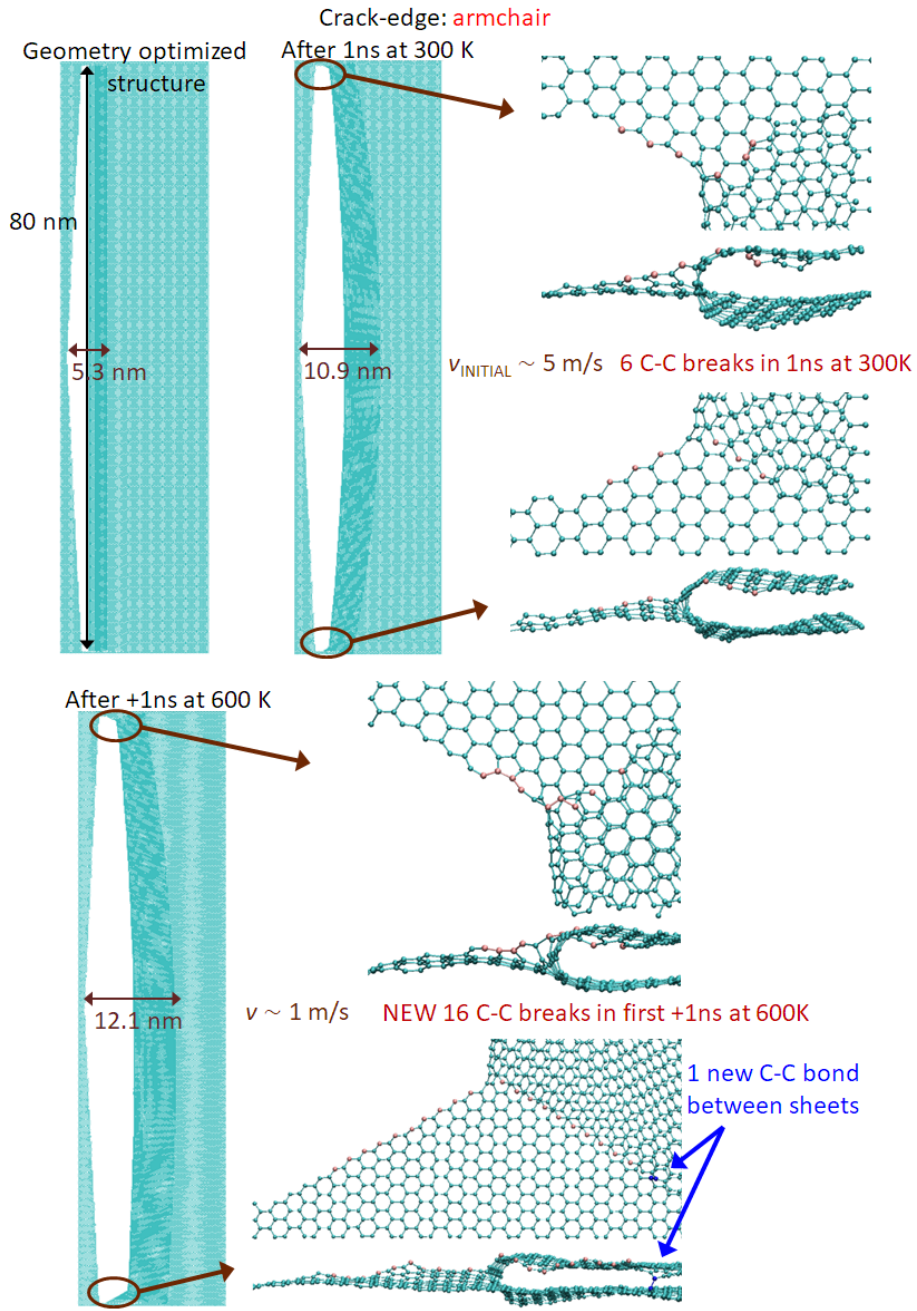

**Figure 3**: Representative MD snapshots of the structure with nanoribbon width $w$ = 80 nm and armchair crack-edge simulated without a substrate. Structures from top to bottom panels: geometry optimized (top left), after 1 ns at 300 K (top right), and after 1ns at 600 K (bottom). The horizontal arrow represents the distance from the vertices of the triangular cut to the nanoribbon front. The estimative of the front velocity is shown along the panels. The second and third panels present insets with magnifications of the local structure of the crack-edge and some additional information on the number of broken C-C bonds during the last 1 ns of simulation. Broken C-C bonds are shown in pink color. C-C bonds between the two adjacent graphene layers are shown in blue color.

In Figure **3**, we present the optimized geometry and the dynamical evolution of a system of $w$ = 80 nm and armchair crack edge simulated without a substrate. The main results for the systems deposited on a substrate are presented in Table **II**. This structure without substrate is shown in order to verify whether nanoribbon growth/propagation would require any attractive support to occur. It also well illustrates the overall form and shape of the crack-propagation. Based on that we did not include additional figures for the same structure for each different substrate. The second panel on top of the Figure **3** shows the structure after 1 ns of simulation at 300 K. The front velocity during this first nanosecond is high (5 m/s) and we observed only a few C-C bonds broke along the line of crack propagation, what is shown in the magnifications of the brown circles (carbon atoms from broken bonds are colored in pink). After an additional period of 1 ns of simulation at 300 K, two new C-C broken bonds along the crack-edge direction were observed and the front velocity was reduced to 0.1 m/s (not shown in Figure **3**). After that, the temperature was increased to 600 K and after an additional 1 ns of MD runs, new 16 C-C broke bonds along the crack edge (bottom panel of Figure **3** and its magnifications) were observed. The front velocity increased to 1 m/s, but before the end of this 1 ns of simulation new C-C bonds are formed between the graphene layers (blue carbon atoms in the magnifications of the third panel of Figure **3**). An additional 4 ns of MD runs at 600 K were, carried out and it was observed (not shown in Figure **3**) the breaking of only 6 additional C-C bonds along the crack edge and the front velocity decreased one order of magnitude to about 0.05 m/s. This indicated that the growth/propagation process was practically stopped, consistent with the fact that no new bonds formed between graphene layers was observed. We would like to emphasize two important features: one is the quite regular zigzag pattern of the broken crack-edge of the bottom part of the structure, as seen in the bottom panel of Figure **3**. The second is that the cracking edges at both sides do not grow/propagates in a synchronized way, i. e., by the same amount at the same time. The additional 6 broken C-C bonds (not shown) occurs later at the upper crack-edge shown in the magnifications of the bottom panel of Figure **3**.

Table **II**: Average front velocity, $v$, number of broken C-C bonds along the crack-edges ($nb$), and number of newly formed C-C bonds between carbon atoms of different graphene layers ($nl$) calculated for every nanosecond of simulation at given values of temperature and strength of adhesion $\varepsilon$, for the structures of width $w \cong 80$ nm. Some rows are shaded in order to facilitate the comparison of the variables against different $\varepsilon$ values.

| Crack-edge type | Zigzag | | | | | | | armchair | | | | | | |
|---|---|---|---|---|---|---|---|---|---|---|---|---|---|---|
| Time [ns] | 1 | 2 | 1 | 2 | 3 | 4 | 5 | 1 | 2 | 1 | 2 | 3 | 4 | 5 |
| Temperature | 300 K | | 600 K | | | | | 300 K | | 600 K | | | | |
| $\varepsilon = 0.0005$ eV | | | | | | | | | | | | | | |
| $v$ [m/s] | 4.30 | 0.27 | 0.24 | 0.18 | 0.26 | 0 | 0 | 4.70 | 0.20 | 1.30 | 0.82 | 0.79 | 0 | 0 |
| $nb$ | 8 | 2 | 7 | 3 | 2 | 1 | 0 | 0 | 3 | 16 | 8 | 2 | 0 | 0 |
| $nl$ | 0 | 0 | 2 | 0 | 0 | 0 | 0 | 0 | 0 | 0 | 0 | 0 | 0 | 0 |
| $\varepsilon = 0.001$ eV | | | | | | | | | | | | | | |
| $v$ [m/s] | 4.30 | 0.27 | 0.42 | 0.29 | 0 | 0 | 0 | 4.80 | 0.15 | 0.98 | 0.37 | 0.44 | 0.02 | 0.13 |
| $nb$ | 3 | 3 | 12 | 1 | 0 | 0 | 0 | 1 | 0 | 14 | 2 | 3 | 5 | 1 |
| $nl$ | 0 | 0 | 0 | 0 | 0 | 0 | 0 | 0 | 0 | 0 | 0 | 0 | 0 | 1 |
| $\varepsilon = 0.003$ eV | | | | | | | | | | | | | | |
| $v$ [m/s] | 4.30 | 0.26 | 0.22 | 0 | 0 | 0 | 0 | 4.80 | 0 | 0.53 | 0.28 | 0.15 | 0 | 0 |
| $nb$ | 5 | 6 | 11 | 0 | 0 | 0 | 0 | 0 | 1 | 13 | 2 | 0 | 0 | 0 |
| $nl$ | 0 | 0 | 2 | 0 | 0 | 0 | 0 | 0 | 0 | 0 | 0 | 0 | 0 | 0 |
| $\varepsilon = 0.005$ eV | | | | | | | | | | | | | | |
| $v$ [m/s] | 4.10 | 0 | 0 | 0.26 | 0.22 | 0 | 0 | 4.60 | 0 | 0.22 | 0 | 0.14 | 0 | 0.24 |
| $nb$ | 1 | 0 | 6 | 2 | 3 | 0 | 0 | 0 | 0 | 5 | 4 | 0 | 0 | 0 |
| $nl$ | 0 | 0 | 1 | 0 | 0 | 0 | 0 | 0 | 0 | 0 | 2 | 0 | 0 | 0 |
| $\varepsilon = 0.010$ eV | | | | | | | | | | | | | | |
| $v$ [m/s] | 3.80 | 0 | 0.48 | 0.24 | 0 | 0 | 0 | 4.20 | 0 | 0.77 | 0.80 | 0 | 0 | 0.28 |
| $nb$ | 0 | 0 | 7 | 5 | 3 | 0 | 0 | 0 | 0 | 8 | 5 | 2 | 0 | 6 |
| $nl$ | 0 | 0 | 0 | 0 | 0 | 0 | 0 | 0 | 0 | 0 | 0 | 0 | 0 | 1 |

In Table **II** we present the data for the number of broken C-C atoms along the crack-edges (*nb*), average front velocity (*v*), and number of newly formed bonds between carbon atoms of different graphene layers (*nl*), for the structures simulated on substrates with different values of the adhesion strength ε (from 0.0005 to 0.01 eV). In order to have a better comparison against the results between structures with armchair and zigzag crack-edges, the values for the zigzag structure in Table **II** were taken from the simulations of a structure with approximately the same size along the direction of the nanoribbon growth/propagation of the armchair structure. The shape in the absence of a substrate and the value of the average front velocity for the case with substrate with adhesion strength ε = 0.01 eV formed the only set of information for this structure reported in Ref. [9].

From Table **II** we can observe that all initial front velocities are larger than 4 m/s. This is an effect of the metastability of the initial geometry optimized configuration. Once the system is subjected to thermal fluctuations, the high adhesion strength between the layers of graphene (~ 0.5 J/m$^2$ [41], almost ten times larger than the adhesion strength between graphene and substrate considered in this study) pulls the nanoribbon fast towards maximizing the formation of bilayer surface. Energy is released that accelerates the nanoribbon. It should be stressed that although these values of the initial front velocity seem high, they are much smaller than the velocities in the natural lattice vibrations of the system that can be estimated by [8] $a_0 * k_B T/h$ ~ 1.42 x 10$^{-10}$ m * 6 x 10$^{12}$ s$^{-1}$ ~ 850 m/s, where $a_0$ is approximately the C-C bond distance, $k_B$ is the Boltzman constant, *T* is the temperature, and *h* is the Planck's constant.

The value of the initial front velocity decreases with the adhesion strength for ε ≥ 0.005 eV or ≥ 0.03 J/m$^2$. Graphite-silica interface, for example, has an adhesion strength of ~ 0.08 J/m$^2$, thus above this minimum value. The approximated rate of decrease of this initial velocity, estimated by the ratio of variation of the front velocity to the variation of the graphene-substrate adhesion strength, is about 12.3 (14.8) m/s per J/m$^2$ for the zigzag (armchair) initial chirality of the crack-edge.

For all cases, the front velocity strongly decreased from the first to the second period of 1 ns of simulation at 300 K. For adhesion strengths larger than 0.003 eV or 0.017 J/m$^2$, the front velocity decreased to practically zero in the second period of 1 ns of simulation. During the first 1 ns of simulation, just after increasing the temperature from 300 to 600 K, the front velocity, in most of cases increased again due to the thermally induced breaking of additional C-C bonds along the crack-edge. However, in some cases, i. e., the values of the front velocity during the first 1 ns at 600 K, were approximately the same as or smaller than the one calculated at previous 1 ns at 300 K. It happened only for the zigzag crack-edge structures with $\varepsilon$ = 0.0005 eV, 0.003 eV and 0.005 eV. Analyzing the data for the number of broken and/or formed C-C bonds, these three cases have in common the fact that new chemical bonds are created involving carbon atoms of adjacent graphene layers. As shown in the insets of Figures **2** and **3**, the number of these bonds can be large enough to prevent or, at least, significantly decrease the nanoribbon growth/propagation. Further MD simulations maintaining the temperature at 600 K showed that the front velocity in the most of the cases reduced up to zero or almost zero.

Another observation that can be inferred from the data presented in Table **II** is about the qualitative relation between the front velocity and number of broken carbon bonds (*nb*) and the number of formed carbon bonds between adjacent graphene layers (*nl*). When the formation of broken bonds is observed, the front velocity increases or, at least, maintains the same value as in the previous period of time. However, sometimes even with non-zero broken carbon bonds, because of the formation of new bonds between the graphene layers, the front velocity decreases. Based on that we can conclude that the front velocity increases with either the increasing temperature, as described in the previous paragraph, or with the breaking of carbon bonds, but decreases with the formation of new bonds between adjacent graphene layers. However, as the increase of temperature led to the increase of the probability of formation of new bonds between the layers, it will also indirectly contribute to the decrease of the front velocity. Thus, although thermal fluctuations can help triggering the break of carbon bonds at the cracking edges, they also can contribute to prevent the nanoribbon growth/propagation process to continue, by increasing the chance of formation of new bonds between graphene layers. From Table **II** we can see that there is no formation of new bonds when the structure is simulated at 300 K. Thus, we could conclude that maintaining the system at room temperature will avoid the undesired formation of bonds between graphene layers. However, in order to verify whether this conclusion is not size-dependent, we also simulated systems with twice the width size, i. e., $w$ = 160 nm.

In Figures **4** and **5**, we present a series of MD snapshots of zigzag and armchair crack-edge structures, on a substrate of adhesion strength of $\varepsilon$ = 0.001 eV or 0.0058 J/m$^2$, for: (i) after geometry optimization; (ii) after 1ns at 300 K; (iii) after 2 ns at 300 K; (iv) after the first 1 ns of simulation at 600K, and; (v) after 6 ns at 300 K. For the structures with $w \cong$ 160 nm, after 2 ns of simulation at 300 K, we performed two additional series of simulations: one with an additional 4 ns of simulation at 300 K (i. e., maintaining the same temperature), and other with more 4 ns of simulation at 600 K. The results are presented and discussed below.

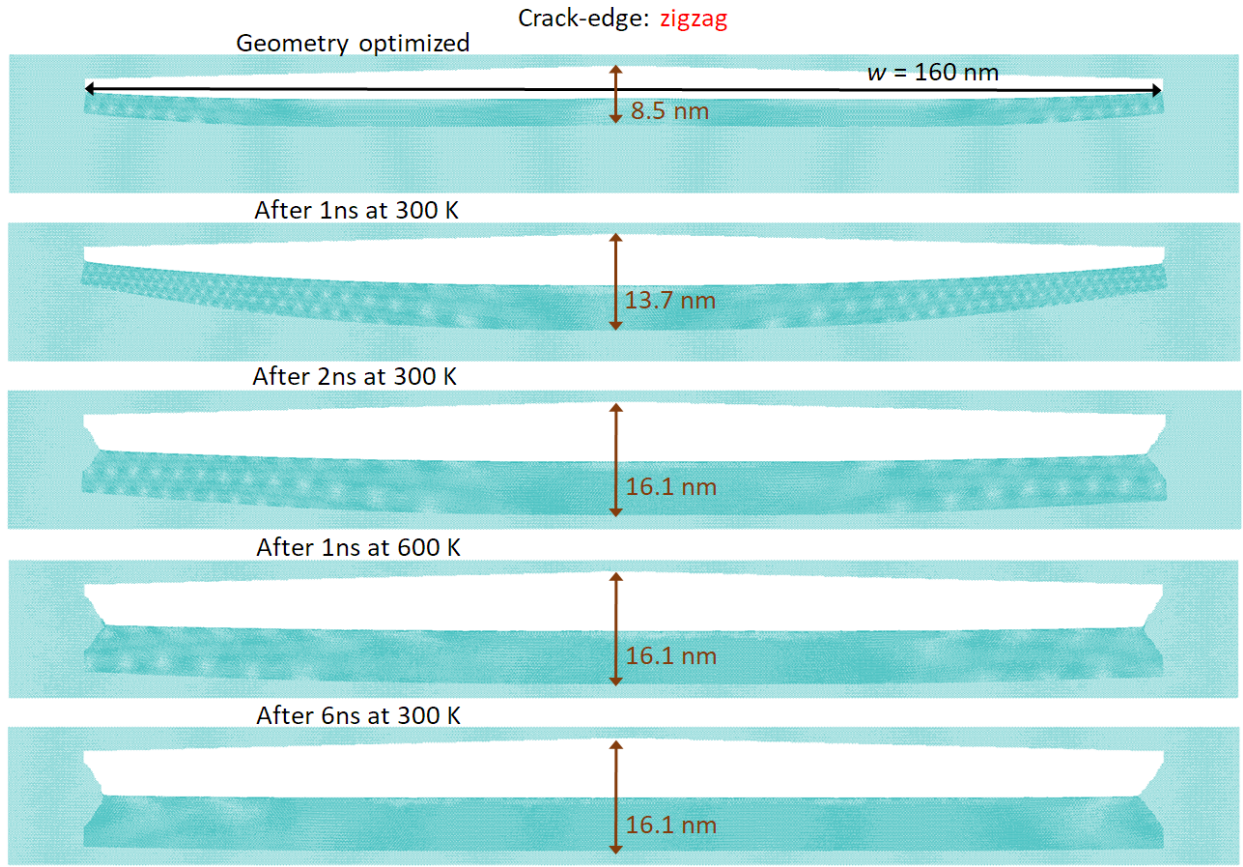

**Figure 4**: MD snapshots of the structure with nanoribbon width *w* = 160 nm and zigzag crack-edge simulated with a substrate of adhesion strength ε = 0.001 eV. Structures from top to bottom panels: geometry optimized, after 1 ns at 300 K, after 2 ns at 300 K, after 1ns at 600 K and after 6 ns at 300 K. The vertical arrows represent the distance from the vertices of the triangular cut to the nanoribbon front.

A visual inspection of Figures **4** and **5** allows the verification of the shape and some structural features along the movement of the nanoribbon. First, the line of the front edge of the graphene nanoribbon became curved, similarly to what was observed for the structures with *w* = 80 nm. The front velocity was also measured using the position of the middle part of the front edge of the nanoribbon. This curvature is the effect of the peeling and tearing resistance forces that act on the crack-edges, which are resistant to the fracture of graphene [8,11]. This also causes the Moiré pattern with an increasing period varying from the edges to the center of the nanoribbon, and disappears when the growth/propagation reaches its maximum. The taper angles, i. e., the angle between the line of nanoribbon growth and the crack-edge line, are almost the same and about 30◦. Although for the initial zigzag crack-edge structure, the crack-edge grows/propagates irregularly, for the initial armchair crack-edge structure, it grows/propagates regularly in a zigzag form (see Fig. **7** below and comments about it).

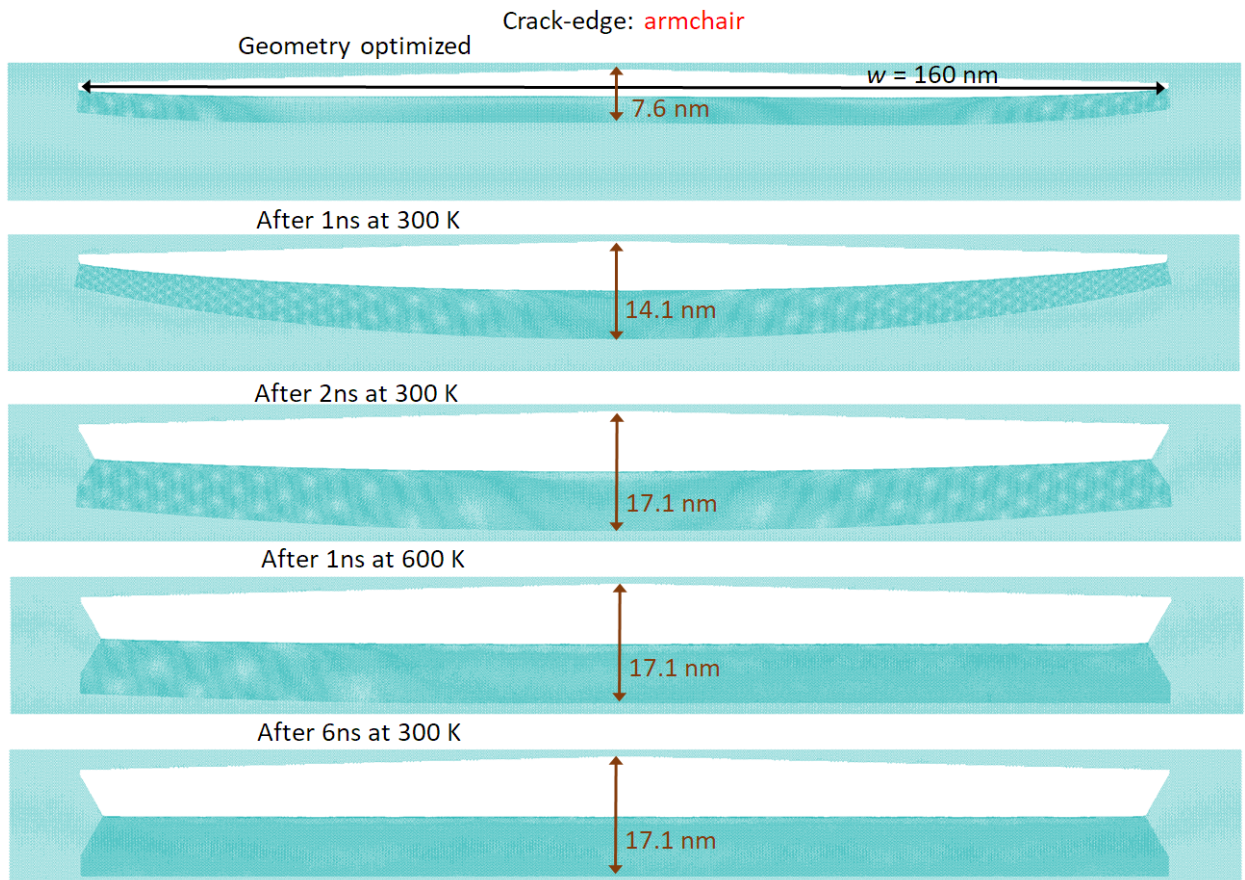

**Figure 5**: MD napshots of the structure with nanoribbon width *w* = 160 nm and armchair crack-edge simulated with a substrate of adhesion strength ε = 0.001 eV. Structures from top to bottom panels: geometry optimized, after 1 ns at 300 K, after 2 ns at 300 K, after 1ns at 600 K and after 6 ns at 300 K. The vertical arrows represent the distance from the vertices of the triangular cut to the nanoribbon front.

The initial front velocities of the structures shown in Figures **4** and **5** are 5.3 m/s and 6.5 m/s for structures with zigzag and armchair initial crack-edges, respectively. The respective numbers of broken C-C bonds at the crack-edge line during the first 1 ns of simulation at 300 K are 4 and 2, while during the second period of 1 ns of simulation these numbers increased to 28 and 35, respectively, with the front velocities being reduced to 2.4 m/s and 3.1 m/s, respectively. Up to 2 ns of simulation at 300 K, no new C-C bonds were formed between the graphene layers. Thus, during the first 2 ns of simulation the most of the C-C bond breaks occurred. We also noticed that the nanoribbon front edge reached its maximum distance to the opposite vertices of the hole during these first 2 ns of simulation. After that, the additional 4 ns of simulations at 300 K showed only additional C-C bond breaks until the front line of the nanoribbon becomes straight (bottom panels of Figures **4** and **5**). However, the additional simulations at 600 K showed that despite the increase of broken C-C bonds, the additional intensity of the thermal fluctuations caused the formation of three new C-C bonds between the graphene layers. In Figure **6**, we present the local structure of one side of the zigzag crack-edge structure, where the carbon atoms forming these new bonds were drawn in dark blue color.

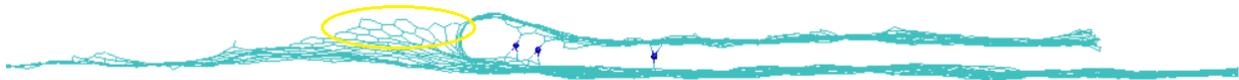

**Figure 6**: Lateral view of the local region of the crack-edge and graphene nanoribbon of the zigzag structure of *w* ≅ 160 nm, after 1 ns of simulation at 600 K, showing the formation of three new C-C bonds between the graphene layers. The atomic structure is drawn in cyan line and the carbon atoms that connect both graphene layers are drawn in dark blue color. The yellow circle highlights the fragment of graphene sheet after local crack-edge formation that is out-of-plane.

Another interesting observation is that the cracking-edge line of the initial armchair crack-edge structure grows/propagates quite regularly and in zigzag form, as can be seen in the right panel of Figure **7**. On the contrary, the cracking-edge line of the initial zigzag crack-edge structure grows/propagates irregularly, exhibiting mixing armchair with zigzag patterns, and presenting some dangling bonds. These patterns are similar to those reported by He *et al*. [16].

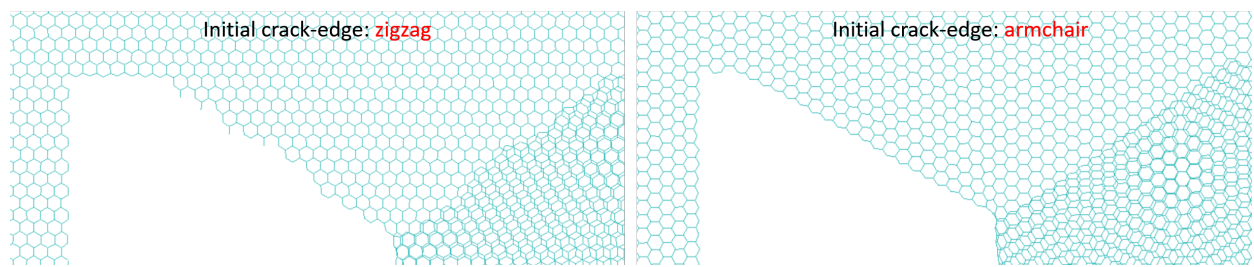

**Figure 7**: Zoomed upper views of the cracking-edge lines grew/propagated after 2 ns at 300 K for the initial zigzag (left) and armchair (right) crack-edge structures.

The yellow circles in Figures **2** and **6** highlight segments of the graphene sheet after local cracking-edge formation due to the nanoribbon growth/propagation. In particular, these segments are spatially out of substrate plane. The orientation and flexibility of these segments decreases the tension on the C-C bonds at the cracking edge point. For example, if we observe the C-C bond on the right side of the yellow region highlighted in Figure **6**, because of the out-of-plane configuration of the neighbor structure, this C-C bond is expected to be less tensioned than if this neighbor structure is closer to the substrate. This type of flexibility of the local structure at the crack-edge could be one of the reasons for decreasing the nanoribbon growth/propagation velocity or even the suppression of the process when the width reaches a critical minimum value.

**4. Conclusions**

We have performed MD simulations of the self-peeling off and self-teared graphene nanoribbon growth on substrates of different adhesion strengths. Structures with initially zigzag and armchair crack-edges were studied and the local atomistic structure of the growing crack-edge as well as some dynamical variables related to the process as the nanoribbon front velocity, number of broken C-C bonds and number of formed bonds between graphene layers were investigated.

In summary, our results show that thermal fluctuations can either favor the process or not, depending on their magnitude. Increase of the temperature leads to the increase of the rate of the breaking of C-C bonds at the cracking-edge. However, increase of the temperature also leads to the increase chance of formation of new C-C bonds between adjacent graphene layers, what helps decreasing the growth process. We observed a direct correlation between the front velocity of the nanoribbon growth and the number of the C-C bond breaking or formation. We found out that the room temperature is a good choice to obtain the best thermal effects on the growth process.

We also observed some features on the atomic structure of the cracking-edges. One of them is the difference in the shape of the cracking-edge line between the structures initially having zigzag and armchair crack-edges. The structure having initial armchair crack-edges gave rise to regular zigzag edges (Figs. **3** and **7**), while initial zigzag crack-edges produced non-regular cracking-edges. Another feature was highlighted in Figs. **2** and **6** by yellow circles. The pieces of the behind remained graphene layer close to the cracking points are not laid on the substrate plane, suggesting that the tension forces at these points might not be as strong as they could if this remained graphene was completely stuck onto substrate. This and the occasional formation of chemical bonds between adjacent graphene layers can contribute to decrease the growth velocity of the graphene nanoribbon.

Our results reveal details of the self-peeling off and self-teared process of formation of graphene nanoribbons on substrates that cannot be observed through the original experiments [8]. However, we believe they can be of help towards the determination of the right conditions to experimentally obtain the formation of nanoribbons of smaller sizes than that reported in [8].

**Acknowledgments**

Authors are fellows the Brazilian Agency CNPq and acknowledge grant #2018/02992-4 from São Paulo Research Foundation (FAPESP), and additional support from the Brazilian Agency CAPES. The authors acknowledge support from the Center for Computational Engineering and Sciences at University of Campinas (FAPESP/CEPID grant No. 2013/08293-7) and from John David Rogers Computing Center (CCJDR) at the Institute of Physics "Gleb Wataghin", University of Campinas.